\pdfoutput=1 
\documentclass{JINST}
\let\ifpdf\relax

\title{uBoost: A boosting method for producing uniform selection efficiencies from multivariate classifiers.}

\author{Justin Stevens and Mike Williams\\
Massachusetts Institute of Technology, Cambridge, MA, United States
}

\abstract{

The use of multivariate classifiers, especially neural networks and decision trees, has become commonplace in particle physics.  Typically, a series of classifiers is trained rather than just one to enhance the performance; this is known as boosting.  This paper presents a novel method of boosting that produces a uniform selection efficiency in a selected multivariate space.  Such a technique is well suited for amplitude analyses or other situations where optimizing a single integrated figure of merit is not what is desired.

}

\begin{document}

\section{Introduction}

Multivariate classifiers are playing an increasingly prominent role in particle physics.  The inclusion of boosted decision trees (BDTs)~\cite{ref:bdt} and artificial neural networks (ANNs) is now commonplace in analysis selection criteria.
BDTs are now even used in software triggers~\cite{ref:lhcbhlt,ref:bbdt}.  It is well known that training a series of classifiers, as opposed to just one, enhances performance.  The training sample for each member of the series is augmented based on the performance of previous members.   Incorrectly classified events are assigned larger weights to boost their importance; this technique is referred to as {\em boosting}.  The result is that each successive classifier is designed to improve the overall performance of the series in the regions where its predecessors have failed.  In the end, the members of the series are combined to produce a single classifier more powerful than any of the individual members.  

The most common usage of BDTs in particle physics is to classify candidates as signal or background.  The structure of the BDT is determined by optimizing a figure of merit (FOM), {\em e.g.}, the approximate signal significance or the Gini index.  This approach is optimal for counting experiments as it by construction produces the optimal selection for observing an excess of events over background.  However, for many types of analyses optimizing an integrated FOM is not what is desired.  For example, in an amplitude analysis obtaining a uniform efficiency in a multivariate space of {\em physics} variates, {\em i.e.}, variates that are of physical interest, is often times more important than any integrated FOM based on the total amount of signal and background.  Such analyses often have many variates in which a uniform efficiency is desired.  A uniform efficiency reduces systematic uncertainties and helps maintain sensitivity to all hypotheses being tested.  

The BDT algorithms available on the market to-date inevitably produce non-uniform selection efficiencies in the variates of physical interest for two reasons: (1)  background events tend to be non-uniformly distributed in the physics variates and (2) the variates provided as input to the BDT have non-uniform discriminating power in the physics variates.
In this paper a novel boosting method, referred to as uBoost, is presented that optimizes an integrated FOM under the constraint that the BDT selection efficiency must be uniform.  The method is described in detail in Section~\ref{sec:method}, while its performance is studied in the context of a Dalitz-plot analysis in Section~\ref{sec:dalitz}.  Some discussion about implementation is provided in Section~\ref{sec:alg} before summarizing in Section~\ref{sec:sum}.  A Dalitz-plot analysis was chosen as the example in this paper due to the simplicity of demonstrating uniformity, or lack thereof, in two dimensions.  The uBoost technique is applicable to situations with any number of variates of physical interest.   

\section{The uBoost Technique}
\label{sec:method}
The variates used in the BDT are denoted by $\vec{x}$ and can be of any dimension.  The variates of physical interest are denoted by $\vec{y}$.  The goal is to produce a uniform selection efficiency in $\vec{y}$; thus, the $\vec{y}$ variates should not also be in $\vec{x}$.  Some subset of $\vec{x}$ are {\em biasing} in $\vec{y}$; {\em i.e.}, their probability density functions (PDFs) vary in $\vec{y}$.  A uniform selection efficiency can, of course, be produced by removing these variates from $\vec{x}$; however, if such a selection does not have adequate discriminating power then the BDT must use biasing variates.  The uBoost algorithm balances the biases to produce the optimal uniform selection.

Boosting works by assigning training events weights based on classification errors made by previous members of the series; events that are misclassified are given more weight.  The uBoost method augments this procedure by not just considering classification error, but also the uniformity of the selection.  Events in regions of $\vec{y}$ where the selection efficiency is lower (higher) than the mean are given larger (smaller) weights.  In this way, uBoost is able to drive the BDT towards a uniform selection efficiency in the $\vec{y}$ variates.  

The uBoost method starts by training the first decision tree (DT) in the standard way: some FOM is chosen, {\em e.g.}, Gini index, and the data is repeatedly split in a way that maximizes this FOM.  For all other trees in the series, events in the training sample are assigned weights based on classification error, denoted by $c$, and non-uniformity, denoted by $u$.  The total weight of event $i$ in tree $t$ is $w^t_i = c^t_i\times u^t_i \times w^{t-1}_i$.
The boosting weight based on classification error, obtained following the AdaBoost~\cite{ref:adaboost} procedure, for event $i$ in the training sample of tree $t$ is given by 
\begin{equation}
        c_i^{t} = \exp{(\alpha_t \gamma_{it})}, 
\end{equation}
where $\alpha_t = \log{((1-e_t)/e_t)}$, $\gamma_{it}$ is one if event $i$ is misclassified in tree $t$ and zero otherwise  and
\begin{equation}
      e_t = \sum_i w^{t-1}_i \gamma_{it}. 
\end{equation}
For each $t$ the $w^t_i$ are normalized such that their sum is unity.  The weights $c_i^{t}$ boost the importance of events that are incorrectly classified by tree $t$.   

The weights based on non-uniformity are designed to boost in importance events in areas of lower-than-average efficiency.  This consideration only applies to signal events; for all background events $u^t_i = 1$.  The $u^t_i$ must be determined independently for all possible mean efficiency values, $\bar{\epsilon}$, since the uniformity of the selection will clearly depend on $\bar{\epsilon}$.  
This means that in principle an infinite number of BDTs is required (one for each $\bar{\epsilon}$ value); however, in practice $\mathcal{O}(100)$ BDTs (1\% steps in $\bar{\epsilon}$) is sufficient.   The CPU cost of uBoost and techniques to reduce the cost are discussed in Sec.~\ref{sec:alg}.

For each value of $\bar{\epsilon}$, the local efficiency for each event, $\epsilon_i^t(\bar{\epsilon})$, is determined using the fraction of the k-nearest-neighbor (kNN) events that pass the BDT cut required to produce a mean efficiency of $\bar{\epsilon}$.  The term BDT here refers to the BDT constructed from the series of DTs trained up to this point in the series.  The $u^t_i$ are then defined as follows:
\begin{equation}
        u^t_i = \exp{(\beta_t (\bar{\epsilon}-\epsilon_i^t))}, 
\end{equation}  
where $\beta_t = \log{((1-e^{\prime}_t)/e^{\prime}_t)}$ and 
\begin{equation}
        e^{\prime}_t = \sum_i w^{t-1}_i c^t_i |\bar{\epsilon}-\epsilon_i^t|.
\end{equation}
 The total event weights are then $w^t_i = c^t_i \times u^t_i \times w^{t-1}_i$.       

The series of DTs trained for any given value of $\bar{\epsilon}$ are combined into a single BDT whose response is given by 
\begin{equation}
   T(\vec{x};\bar{\epsilon}) = \sum_t \alpha_t T_t(\vec{x};\bar{\epsilon}),
\end{equation}
where $T_t(\vec{x};\bar{\epsilon})$ is the response of tree $t$ in the $\bar{\epsilon}$ series.   $T_t(\vec{x};\bar{\epsilon})$ is one if $\vec{x}$ resides on a signal leaf and minus one otherwise\footnote{Another option is to use the purity of the leaf.  This can also be used in conjunction with uBoost.}.  
Each $T(\vec{x}; \bar{\epsilon})$ has a {\em proper} response value associated with it, $\overline{T}(\bar{\epsilon})$, such that the fraction of signal events with $T(\vec{x}_i; \bar{\epsilon}) > \overline{T}(\bar{\epsilon})$ is $\bar{\epsilon}$.
Fortunately, the analyst does not need to see this level of complexity.   Instead, the analyst simply sees a single BDT whose response is
\begin{equation}
\mathcal{T}(\vec{x}) = \frac{1}{N} \sum \Theta(T(\vec{x}; \bar{\epsilon}) - \overline{T}(\bar{\epsilon})),
\end{equation}
where $\Theta$ is the Heaviside function, $N$ is the number of $\bar{\epsilon}$ values used; {\em i.e.},
$\mathcal{T}(\vec{x})$, is the fraction of $\bar{\epsilon}$ values for which $T(\vec{x}; \bar{\epsilon}) > \overline{T}(\bar{\epsilon})$.  
A cut of $\mathcal{T}(\vec{x}) > \bar{\epsilon}$ produces a selection whose mean efficiency is approximately $(1-\bar{\epsilon})$.   
This is the so-called uBoost DT (uBDT).

The value of $k$, the number of nearest-neighbor events used to determine the local efficiency, is a free parameter in uBoost.  The value of $k$ should be chosen to be small enough such that BDT signal selection efficiency is approximately uniform within all hyper-spheres used to collect the kNN events.  However, if $k$ is chosen to be too small, the statistical uncertainty on the local efficiency will be large reducing the effectiveness of uBoost.  This dictates choosing $k$ to be $\mathcal{O}(100)$.  Our studies show very little difference in performance for values of $k$ ranging from 50 to 1000, but a reduction in uniformity for $k < 20$.   The value $k=100$ should be considered as the default. 

\section{Performance of uBoost in Toy-Model Analyses}
\label{sec:dalitz}
 
\begin{figure}
	\begin{center}
		\includegraphics[width=0.49\textwidth]{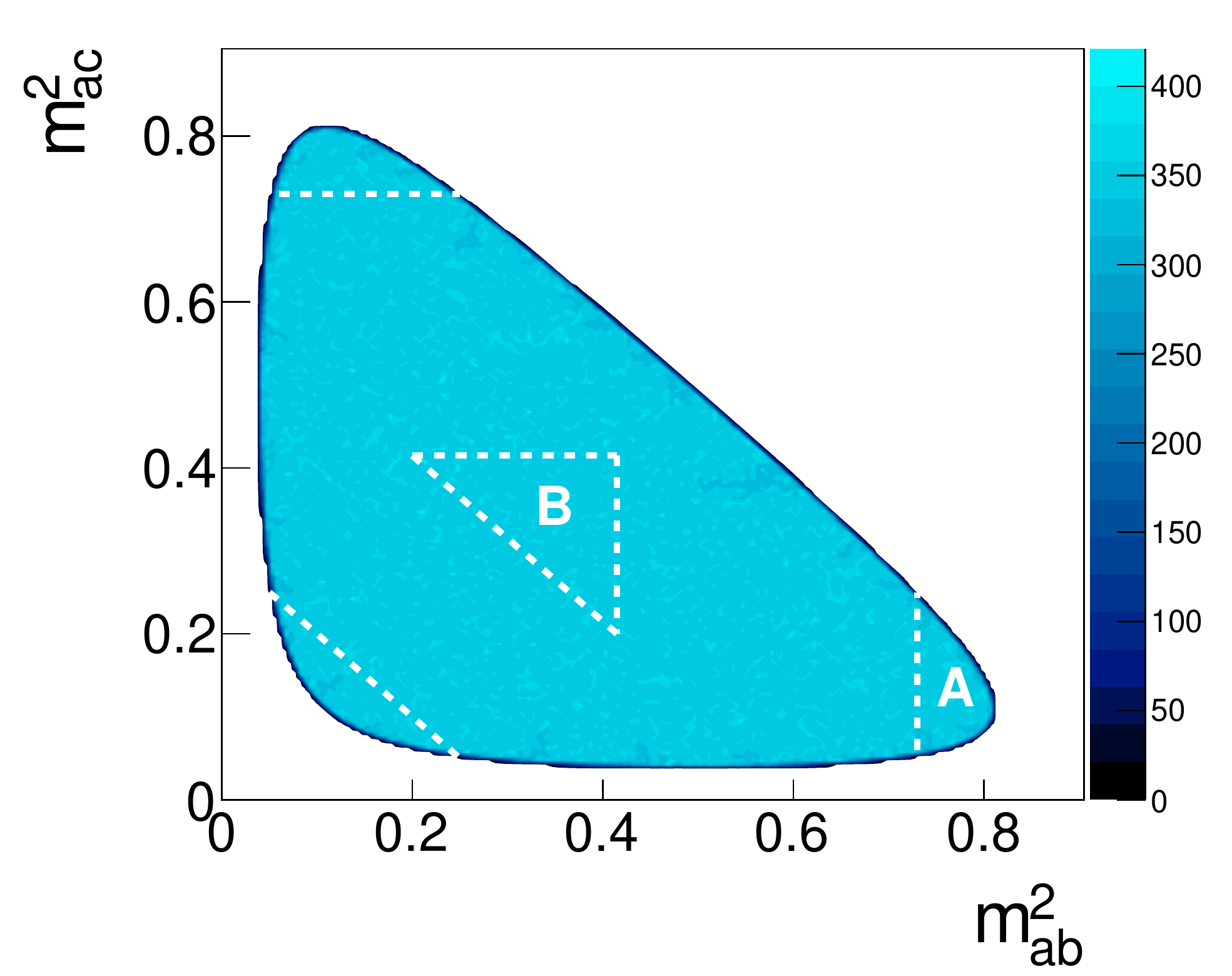} 
		\includegraphics[width=0.49\textwidth]{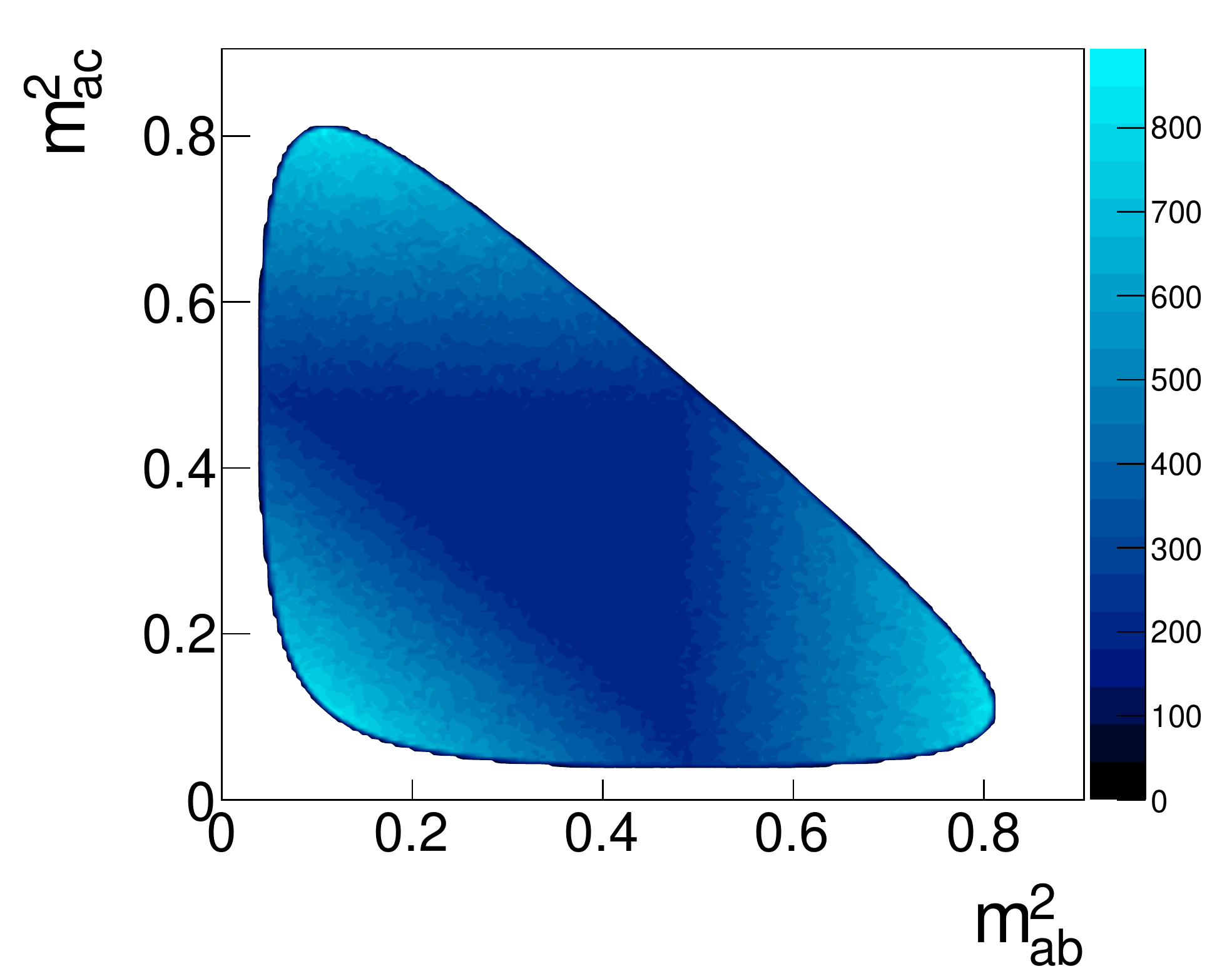}
		\caption{ Dalitz-plot distributions for (left) uniform signal and (right) non-uniform background.  The corner regions, Region A, and central region, Region B, are labeled on the left plot.}
		\label{fig:model}
	\end{center}
\end{figure}

\begin{figure}
	\begin{center}
		\includegraphics[width=0.32\textwidth]{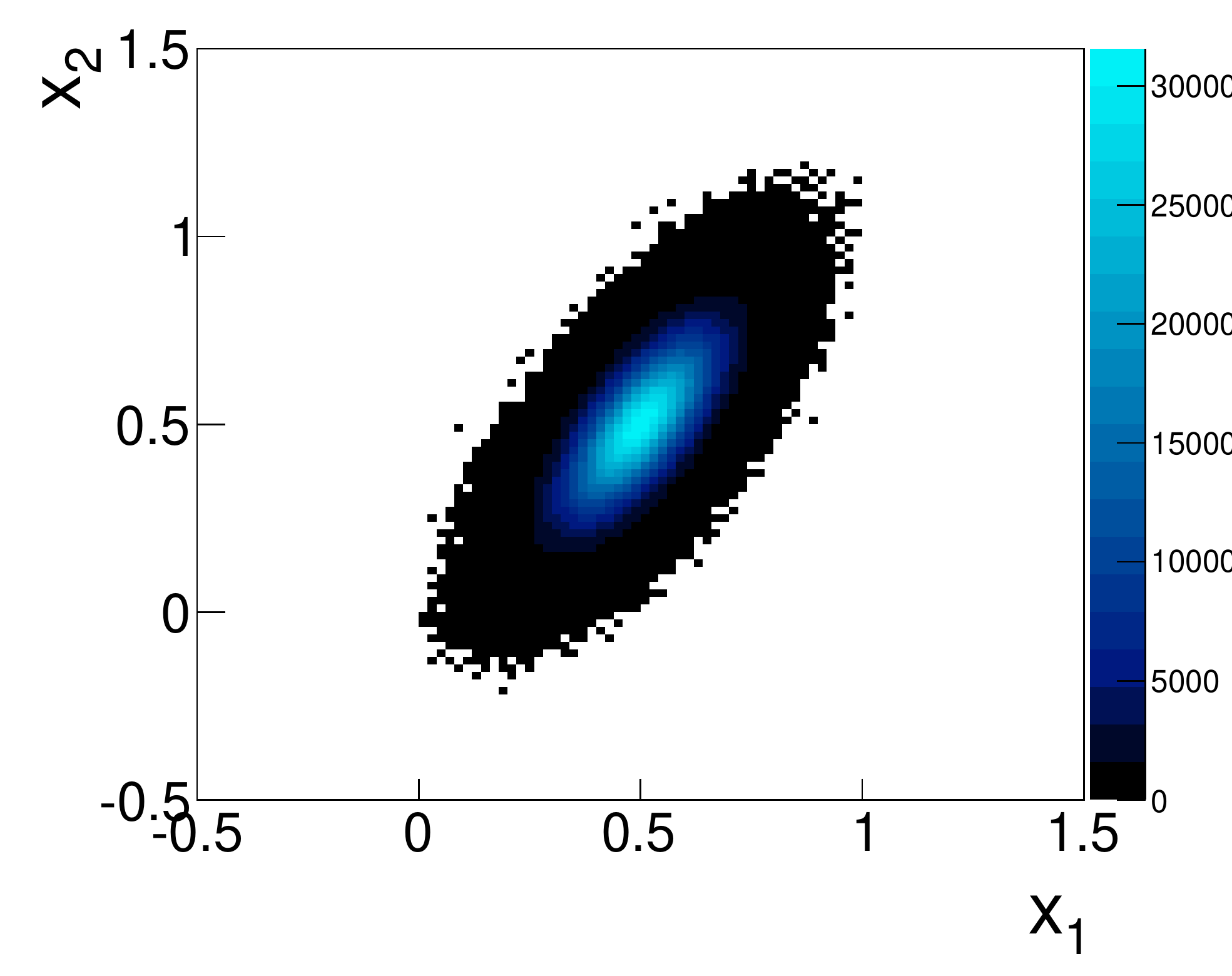}
		\includegraphics[width=0.32\textwidth]{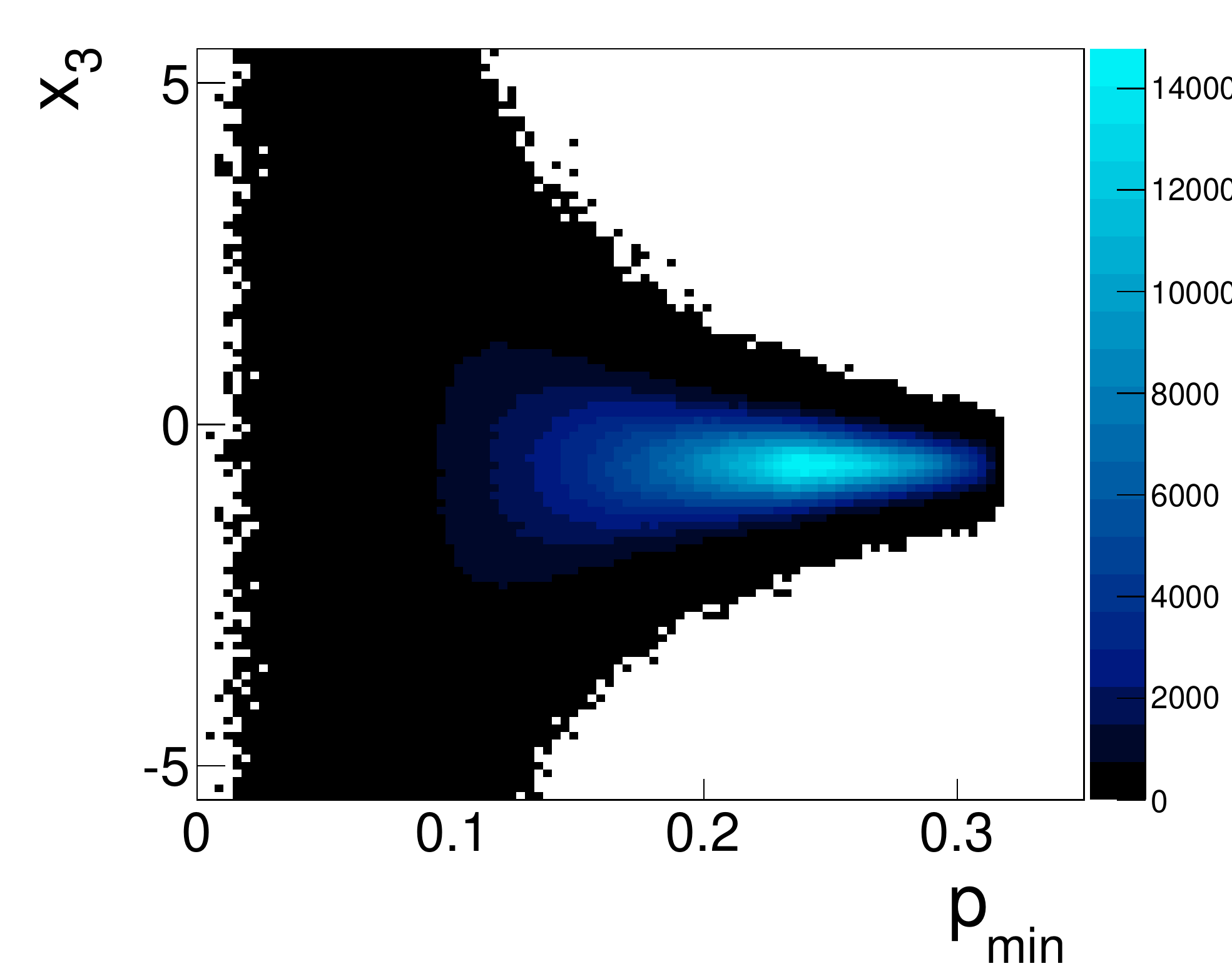} 
		\includegraphics[width=0.32\textwidth]{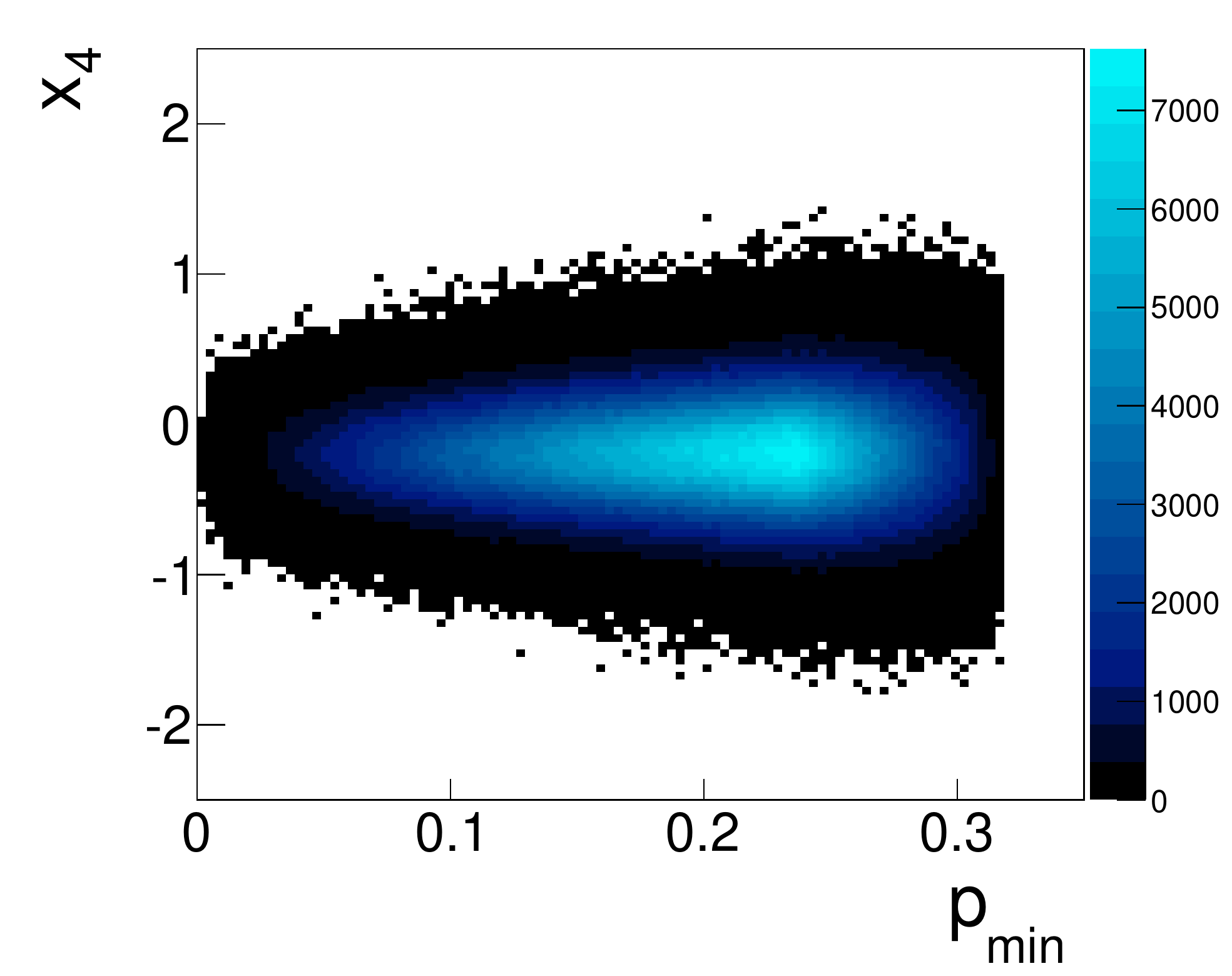}\\
		\includegraphics[width=0.32\textwidth]{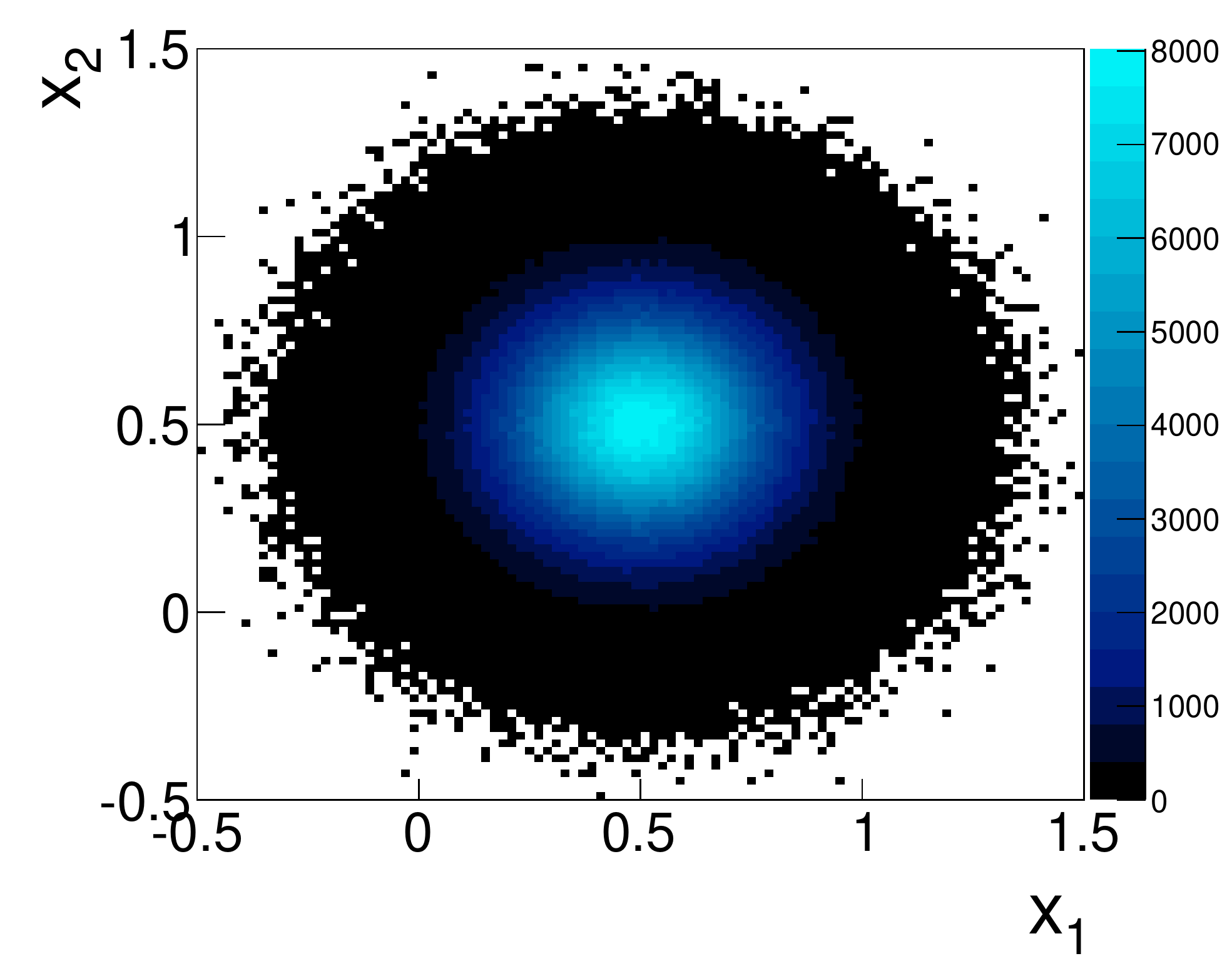}
		\includegraphics[width=0.32\textwidth]{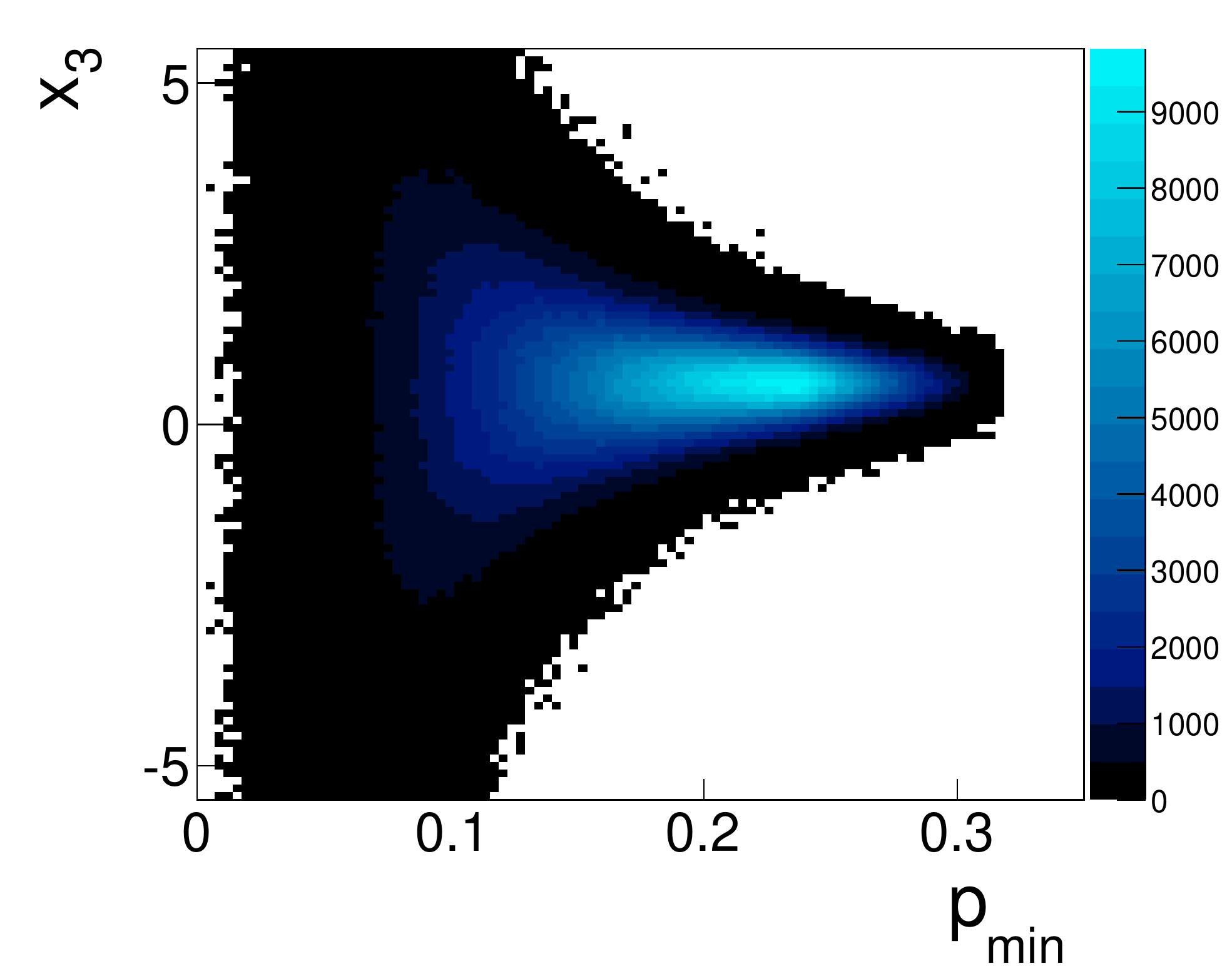} 
		\includegraphics[width=0.32\textwidth]{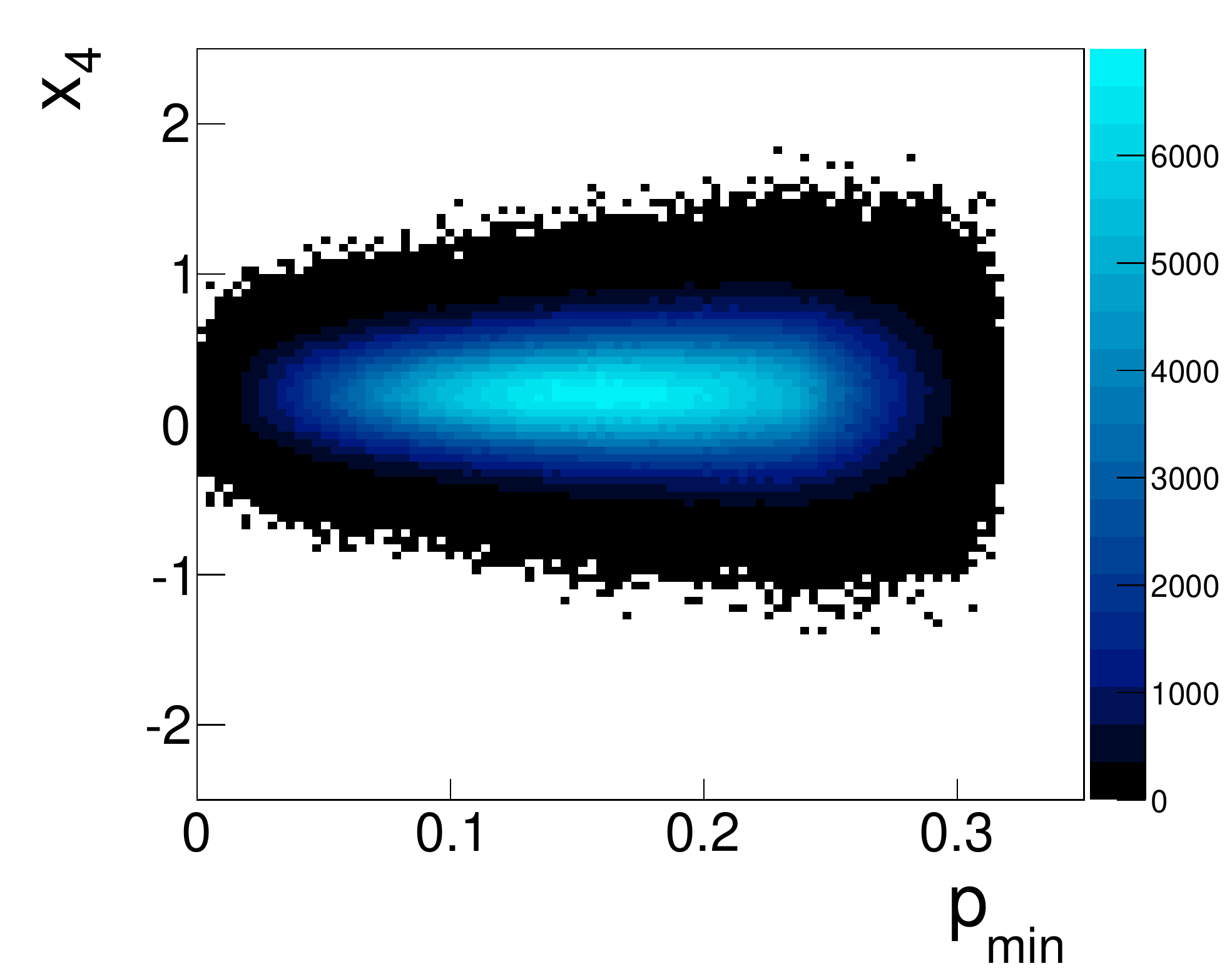}
		\caption{$\vec{x}$ variates: (left) $x_1$ vs $x_2$; (middle) $x_3$ vs $p_{\rm min}$ and (right) $x_4$ vs $p_{\rm min}$ for signal (top) and Model I background (bottom).}
		\label{fig:vars}
	\end{center}
\end{figure}

A Dalitz-plot analysis is used to demonstrate how uBoost works.  The fictional decay $X \to abc$ is considered where $m_X = 1$ and $m_a = m_b = m_c = 0.1$ are the particle masses (in some units) and the minimum momentum of $a$, $b$, and $c$ is denoted $p_{\rm min}$.  The Dalitz variates are given by $\vec{y} = (m^2_{ab},m^2_{ac})$.
Signal events are generated uniformly in $\vec{y}$, while background events either favor regions where one particle has very low momentum (see Fig.~\ref{fig:model}) or are uniform (different models are considered below).  
In the models discussed in this section, we will consider two $\vec{x}$ variates which are biasing in $\vec{y}$ and two which are not.  The variates $x_1$ and $x_2$ are uncorrelated with $\vec{y}$ and are therefore not biasing; however, it is assumed that for this fictional analysis, these two variates alone do not provide sufficient discriminating power.  The two biasing variates, $x_3$ and $x_4$, are shown in Fig.~\ref{fig:vars} as a function of $p_{\rm  min}$ for the signal and background, respectively.  The variate $x_3$ provides strong discrimination for high $p_{\rm  min}$ and deteriorates at low $p_{min}$, while the converse is true for $x_4$.  The variate $p_{\rm min}$ is directly kinematically related to the Dalitz variates $\vec{y}$ and so will not be considered for inclusion in $\vec{x}$.  The model variates are summarized in Table~\ref{tab:model}.

\begin{table}[t]
  \caption{Summary of variates used in the toy Dalitz-plot model.}
\begin{center}\begin{tabular}{|c|c|}
    \hline
variate & comment\\
\hline
$x_1$ & only correlated with $x_2$ (not biasing) \\
$x_2$ & only correlated with $x_1$ (not biasing) \\
$x_3$ & correlated with $p_{\rm min}$ (biasing); more powerful at center of Dalitz plot \\
$x_4$ & correlated with $p_{\rm min}$ (biasing); more powerful at corners of Dalitz plot \\
\hline
\end{tabular}\end{center}
\label{tab:model}
\end{table}

\begin{figure}
	\begin{center}
		\includegraphics[width=0.49\textwidth]{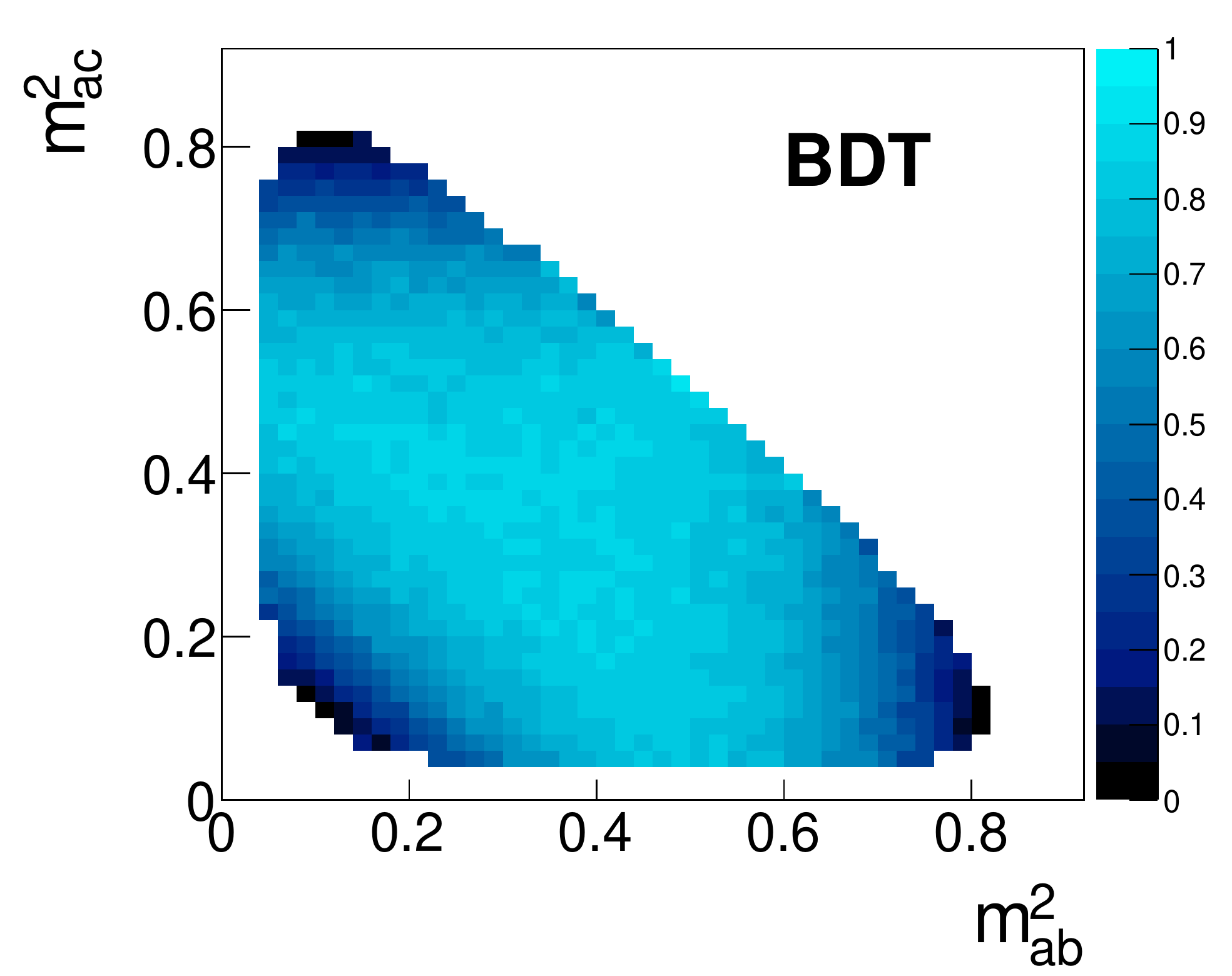} 
		\includegraphics[width=0.49\textwidth]{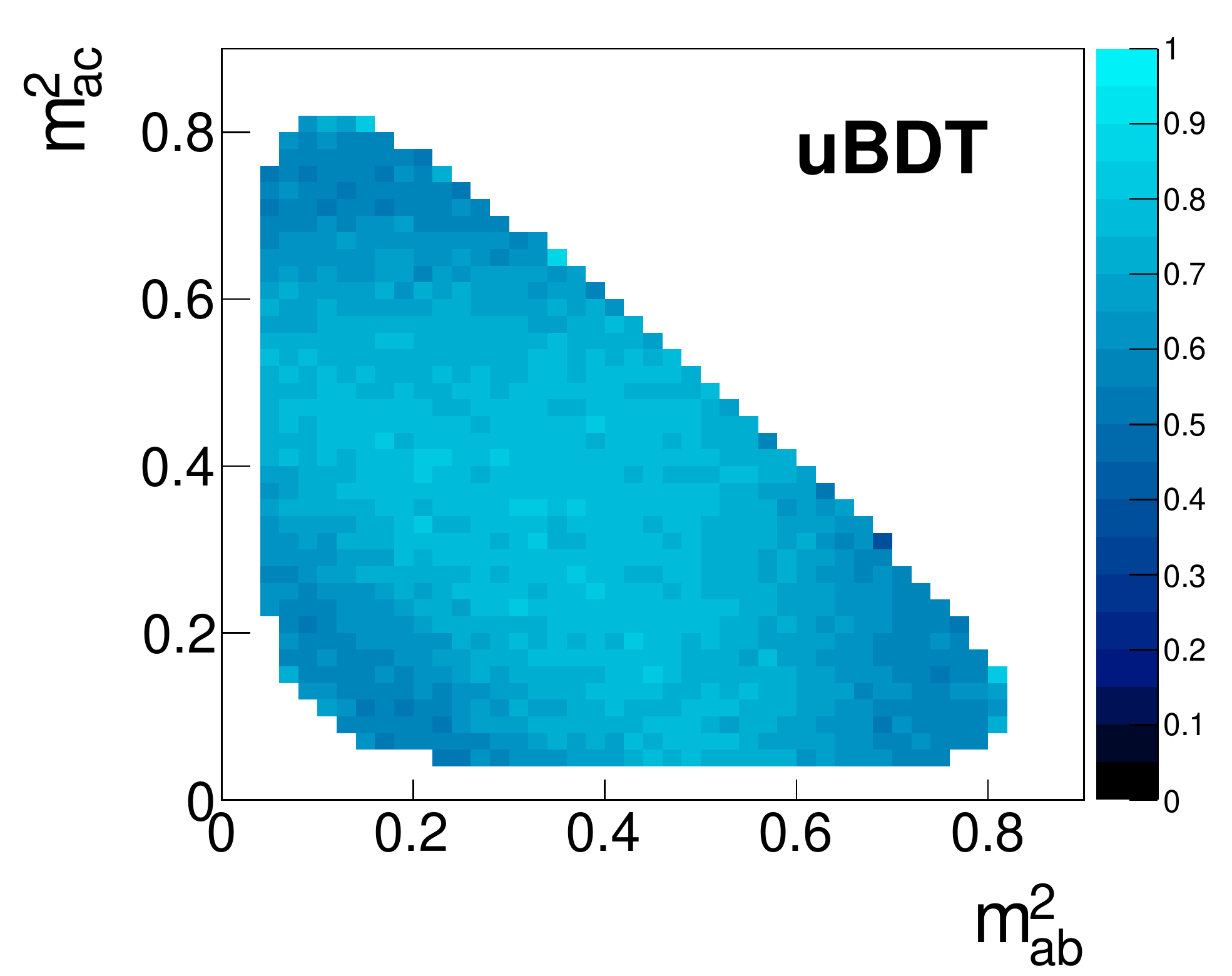}
		\caption{ Model I selection efficiency distributions for $\bar{\epsilon} = 70\%$ from (left) AdaBoost and (right) uBoost.}
		\label{fig:dalitzEffic1}
	\end{center}
\end{figure}

\begin{figure}
	\begin{center}
		\includegraphics[width=0.49\textwidth]{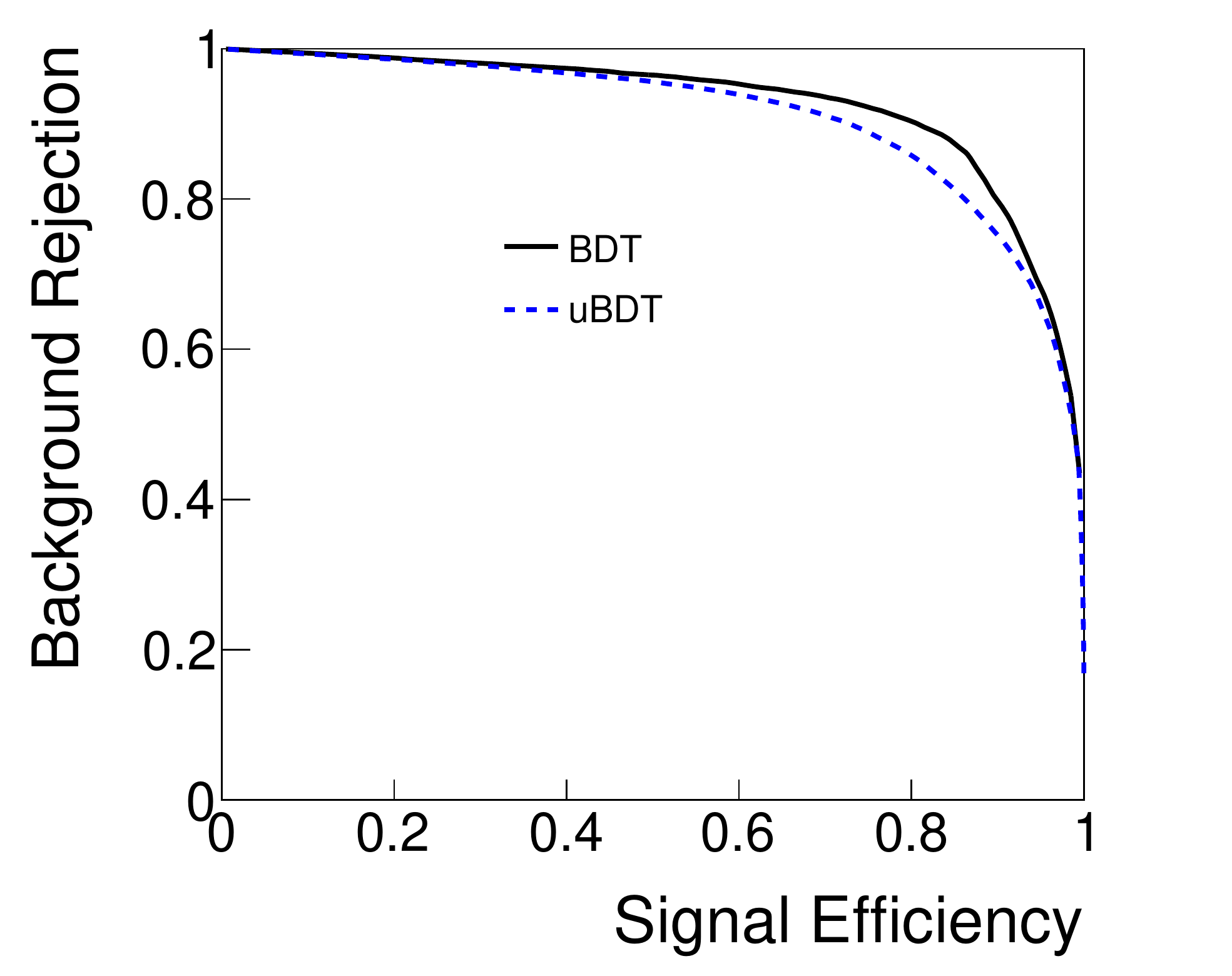} 
		\includegraphics[width=0.49\textwidth]{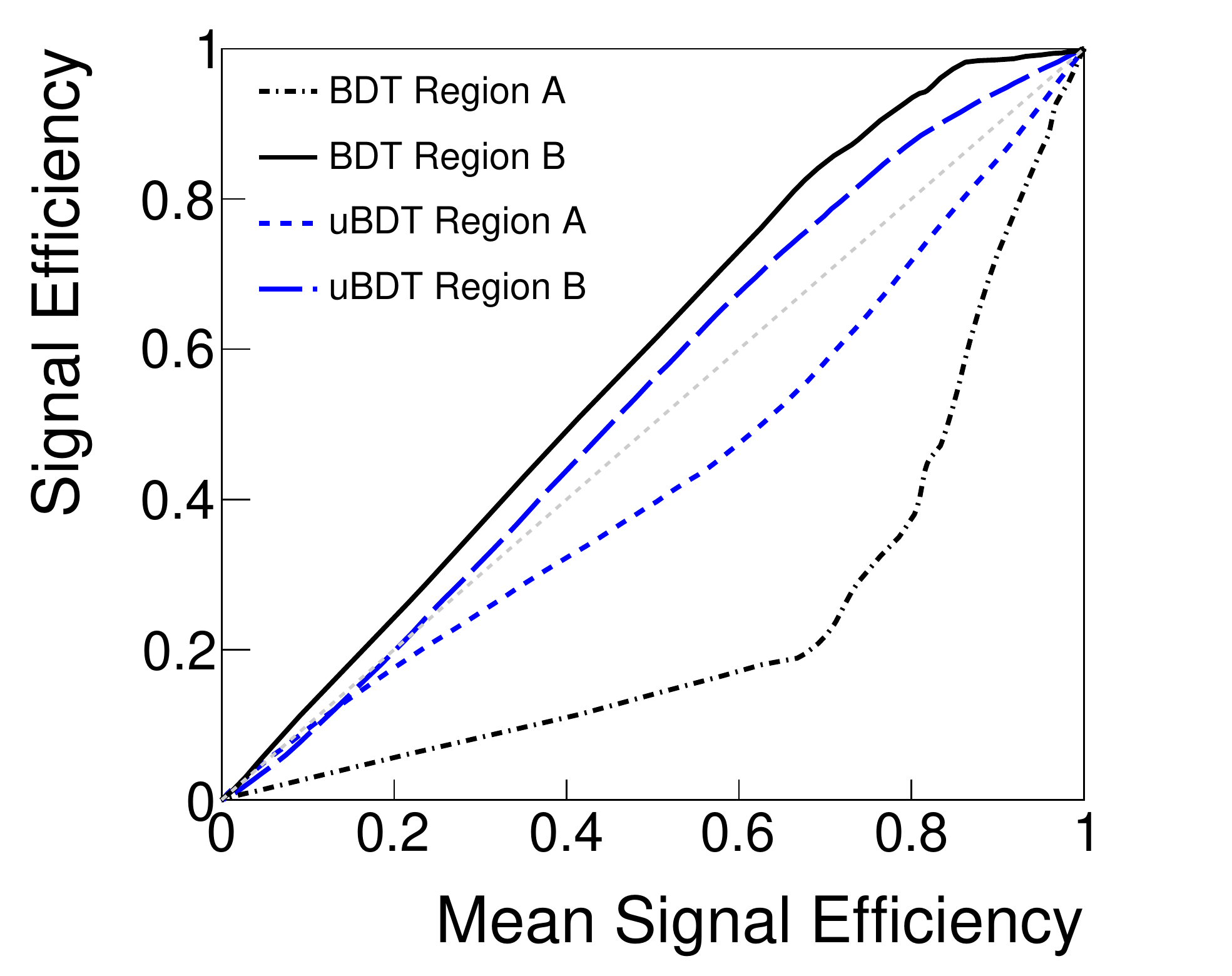}
		\caption{(left) ROC curves and (right) signal efficiencies {\em vs} $\bar{\epsilon}$ in regions A and B for AdaBoost and uBoost for Model I. }
		\label{fig:roc}
	\end{center}
\end{figure}

The first model (Model I) considered utilizes only variates $\vec{x} = (x_1,x_2,x_3)$ in the BDT training and selection process.  For Model I, the non-uniform background distribution shown in Fig.~\ref{fig:model} is used.
Both the distribution of background events and the worse resolution of $x_3$ at low $p_{\rm min}$ contribute to a bias in the standard BDT selection efficiency.   Figure~\ref{fig:dalitzEffic1} (left) shows that the BDT efficiency (for an arbitrary choice of $\bar{\epsilon} = 70\%$) is much lower at the corners of the Dalitz plot than in the center.  This is not a pathology; it is the optimal selection given the variates provided as input to the BDT training and the defined task of optimizing an integrated FOM.  
Figure~\ref{fig:roc}(right) shows the efficiency in the center (Region A) and corners (Region B) of the Dalitz plot as a function of the mean efficiency (Regions A and B are shown on Fig.~\ref{fig:model}(left)).  There is a sharp drop in efficiency at the corners.  
Typically, in a Dalitz-plot analysis the most interesting regions physically are the edges since these areas contain most of the resonance contributions; thus, the BDT selection obtained here is undesirable.  

Figures~\ref{fig:dalitzEffic1}(right) and \ref{fig:roc}(right) show the results obtained for Model I using uBoost where the Dalitz variates $\vec{y}$ are chosen as those of physical interest.   The selection efficiency is now only weakly dependent on $\vec{y}$ as desired.  As expected, the ROC curve shown in Fig.~\ref{fig:roc}(left) reveals that the performance in the integrated FOMs is reduced; however, for this analysis a small reduction in the integrated FOM is acceptable given the large gain in efficiency at the corners of the Dalitz plot.  
In the context of this analysis, the performance of the uBDT is much better than that of the BDT.

\begin{figure}
	\begin{center}
		\includegraphics[width=0.49\textwidth]{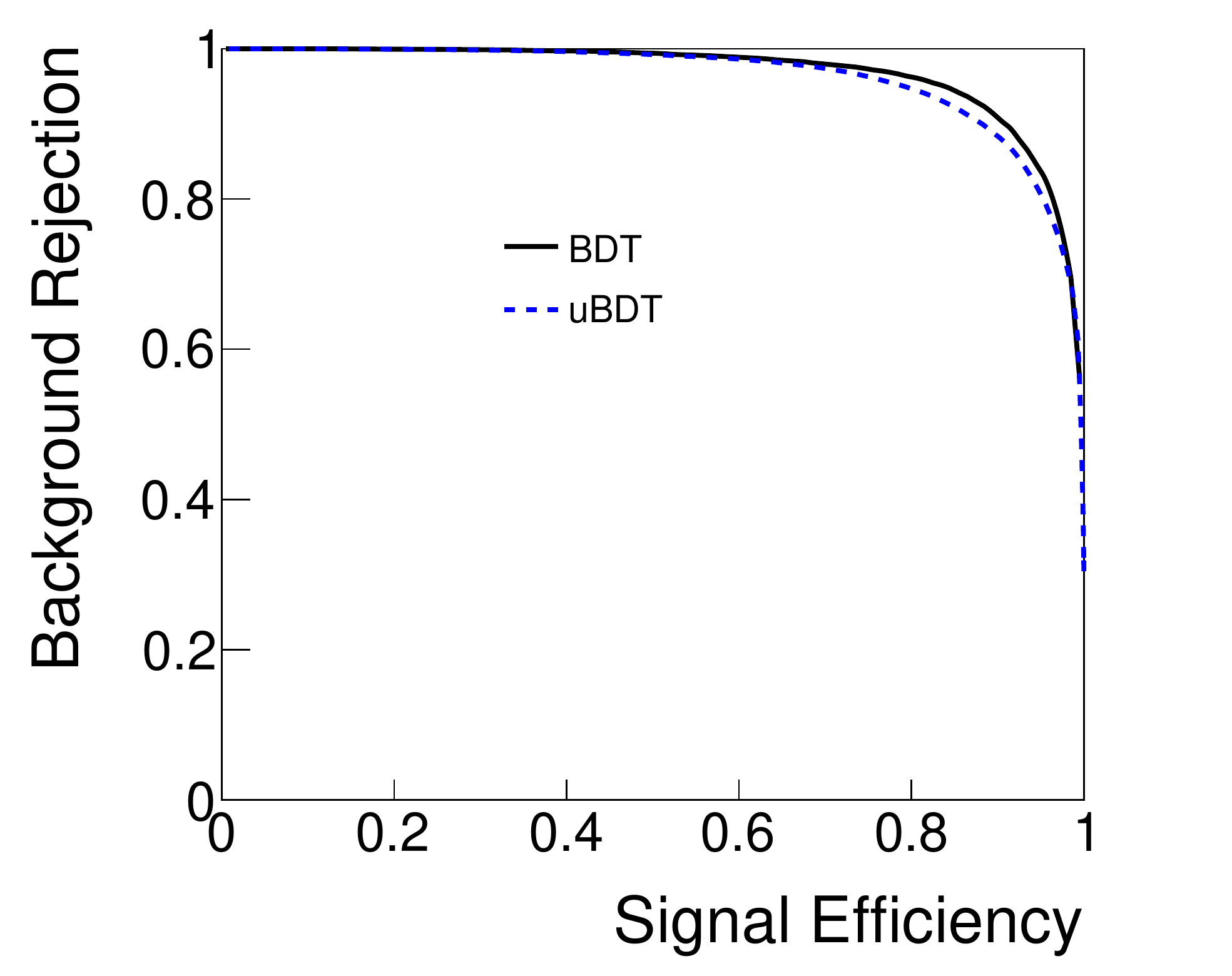} 
		\includegraphics[width=0.49\textwidth]{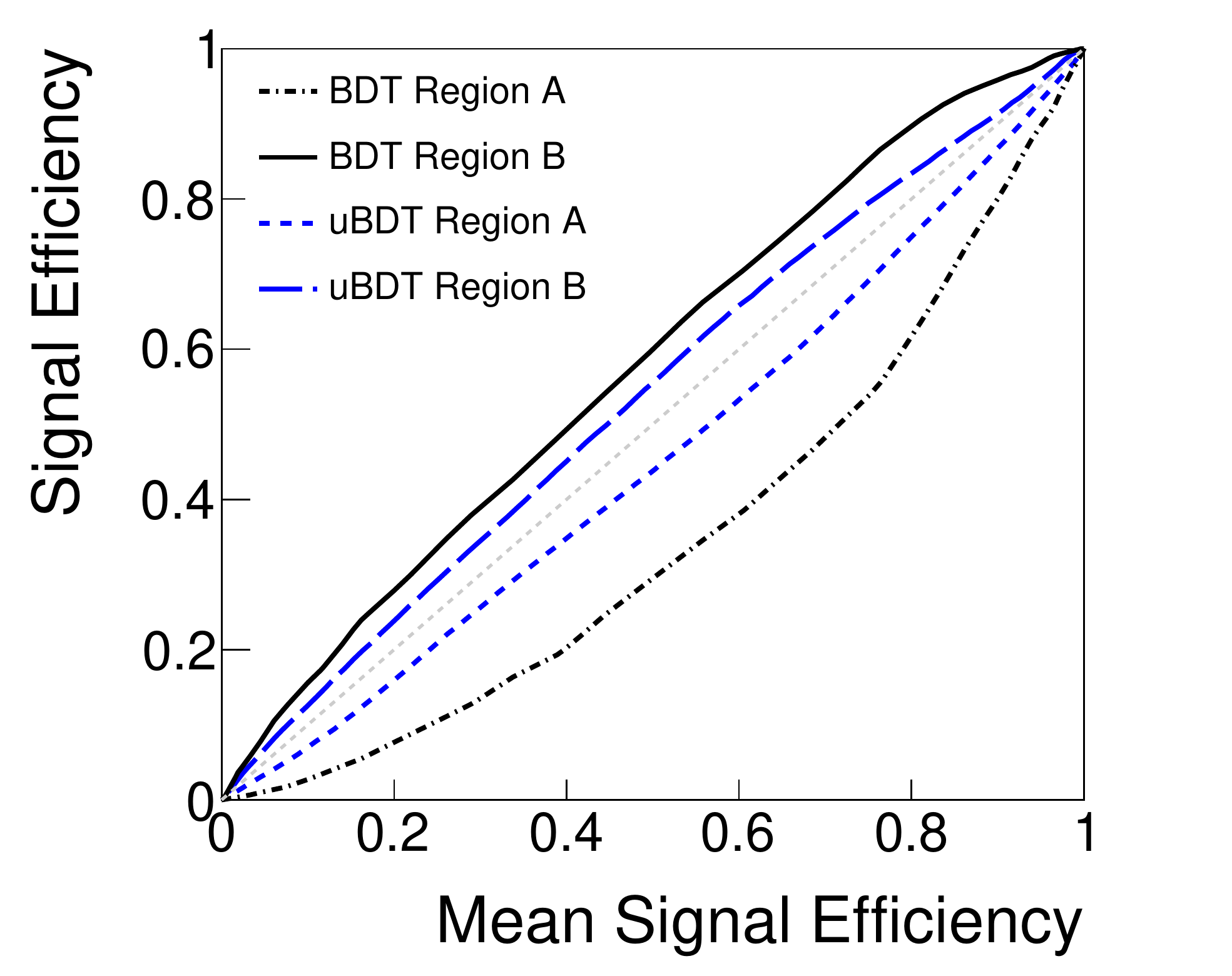}
		\caption{(left) ROC curves and (right) signal efficiencies {\em vs} $\bar{\epsilon}$ in regions A and B for AdaBoost and uBoost for Model II. }
		\label{fig:roc2}
	\end{center}
\end{figure}

Another model (Model II) is considered where both the signal and background are uniformly distributed in $\vec{y}$.  Here, all of the bias must come from differences in the distributions of $\vec{x}$ for signal and background.  For Model II, the input variates to the BDT training are $\vec{x} = (x_1,x_2,x_3,x_4)$; thus, in this case $\vec{x}$ contains a variate that bias towards the center as before but also one, $x_4$, that biases towards the corners.  
Figure~\ref{fig:roc2} shows that for this case, as expected, the BDT is not as biasing as for Model I but the uBDT is still able to produce a selection with much less dependence on $\vec{y}$.  Furthermore, since the uBDT has access to variates that bias the selection in both directions it is able to more effectively balance them and produce a ROC curve that is nearly identical to that of the BDT.
In both models studied, the uBoost method trades a small amount of performance in the integrated FOMs for a large decrease in the dependence of the selection efficiency on $\vec{y}$.  For a Dalitz-plot analysis, this is a desirable trade.  



For this work, AdaBoost was chosen to determine the misclassification weights and for how to combine the DTs in a series into a BDT.   Other methods could be substituted here.    For example, a number of bootstrap-copy data sets could be produced and individual uBDTs trained on each one.  These could then be combined to produce a single uBDT as in the {\em bagging} method~\cite{ref:bag}.  Within a software framework that already supports these techniques implementation would be trivial allowing analysts to easily study multiple configurations of uBoost and choose the one that works best for their analysis.

We conclude this section by discussing an alternative method to uBoost.  For this example analysis, one can divide up the Dalitz-plot into bins and train an independent BDT in each bin.  Uniform efficiency is then obtained by determining the cuts on each BDT response that produce a given target efficiency.   As a benchmark, we have developed such a selection using 100 bins in the Dalitz plot.  In each bin an independent BDT was trained.  Some care was required in defining a binning scheme that does not produce any low-occupancy bins near the edges of the Dalitz plot.  The background rejection obtained with this method is nearly identical to that of uBoost (they are consistent within statistical uncertainties); however, there are disadvantages to using this {\em brute force} approach: it scales poorly to a higher number of variates of physical interest; it produces a discontinuous selection and it requires the definition of a binning scheme with adequate occupancy in each bin.   The number of bins, or BDTs, increases exponentially with the number of variates of physical interest.  Thus, in higher dimensions uBoost is the only practical option, while in a few dimensions the advantages of uBoost still make it the more desirable choice.

\section{Discussion on Computational Performance}
\label{sec:alg}

The BDT training and selection used in this study were implemented within the framework of the Toolkit for Multivariate Data Analysis (TMVA)~\cite{ref:TMVA2007} using a private copy of the TMVA source code.
The uBoost method as described in Sec.~\ref{sec:method} requires more CPU resources than AdaBoost.  The training time scales up by the number of $\bar{\epsilon}$ values for which a BDT is trained which is of $\mathcal{O}(100)$.  If this CPU price is too steep, then a smaller number of $\bar{\epsilon}$ values can be used resulting in a decrease in the ROC performance.  Another CPU cost is incurred when determining the non-uniformity-based event weights which requires evaluating the BDT (at the current point in the training series) once for each training signal event (for each DT trained). There is also a CPU price to pay for determining the k-nearest neighbors for each event; however, this only needs to be done once prior to training the uBDT and many algorithms exist (see, {\em e.g.}, Ref.~\cite{ref:knn}; we used the $k$-d tree algorithm implemented in TMVA) that make this take a negligible amount of time.  The response time is increased due to the fact that there are more BDTs (one for each $\bar{\epsilon}$ value) that need to be evaluated.  The average size of the BDTs is also increased with respect to AdaBoost.  In AdaBoost, trees near the end of the series have fewer leaves since fewer events carry a larger fraction of the weight.  This does not happen in uBoost leading to, on average, more leaves (about 50\% more in our studies).  

For the toy model studied in this paper, the training signal and background samples contained 25 thousand events each.  
For each value of $\bar{\epsilon}$, a series of 100 DTs was trained and $k=100$ was chosen for the nearest-neighbor algorithm.  
BDTs were trained for 100 values of $\bar{\epsilon}$: 0.01 to 1.0 in step sizes of 0.01. The training time for each of the uBoost BDTs for each of the 100 $\bar{\epsilon}$ values was about 25\% larger than for a single AdaBoost DT.   The total uBDT training time was $100 \times 1.25 = 125$ times more than for a standard BDT.   For these parameters, the standard BDT took a few seconds to train on a single CPU core while the uBDT took several minutes.   Of course, if the analyst has {\em a priori} knowledge of the efficiency range of interest then the range of $\bar{\epsilon}$ values can be restricted to this range which would decrease the CPU time required.  
The training time scales linearly with the number of trees trained and with the number of input variates; it scales as $N\log{N}$ for $N$ total training events.  Thus, the training times should not be prohibitive.  Furthermore, the uBDT method is easily parallelized since the BDTs for each $\bar{\epsilon}$ value are independent.  This permits reduction in the real time cost by increasing the number of CPU cores.  
Finally, the response in our study was evaluated at close to 20~kHz for AdaBoost compared to 250~Hz for uBoost.  This scaling is close to the number of $\bar{\epsilon}$ values chosen.

\section{Summary}
\label{sec:sum}
 
A novel boosting procedure, uBoost, has been presented that considers the uniformity of the selection efficiency in a multivariate space in addition to the traditional misclassification errors.  
 The algorithm requires more CPU time than traditional BDTs but not a prohibitive amount more.  
The uBoost method is expected to be useful for any analysis where uniformity in selection efficiency is desired, {\em e.g.},  in an amplitude analysis. 
\acknowledgments

This work was supported by the US Department of Energy (DOE) under cooperative research grant DE-FG02-94ER-40818.

\end{document}